\RequirePackage{fix-cm}
\documentclass[twocolumn]{svjour3}          
\smartqed  
\usepackage{graphicx}

\newcommand{\caldot}[1]{{\dot{\mathcal{#1}}}}
\begin{document}

\title{Efficiency of inefficient endoreversible thermal machines
}

\subtitle{}


\author{Jos\'e P. Palao, Luis A. Correa, Gerardo Adesso and Daniel Alonso}


\institute{Jos\'e P. Palao \at IUdEA and Department of Physics, University of La Laguna, La Laguna 38200, Spain 
\and
Luis A. Correa \at Department of Physics, Universidad Aut\'onoma de Barcelona - E08193 Bellaterra, Spain 
\and
Gerardo Addesso \at School of Mathematical Sciences, University of Nottingham, University Park, Nottingham NG7 2RD, UK
\and
Daniel Alonso \at IUdEA and Department of Physics, University of La Laguna, La Laguna 38200, Spain \\
\email{dalonso@ull.es}   
}

\date{Received: date / Accepted: date}

\maketitle
\begin{abstract}

We present a study of the performance of endoreversible thermal machines optimized with respect to the thermodynamic force associated with the cold bath in the regime of small thermodynamic forces. These thermal machines can work either as an engine or as a refrigerator.  We analyze how  the optimal performances are determined by the dependence of the thermodynamic flux on the forces. The results are motivated and illustrated with a quantum model, the three level maser, and explicit analytical expressions of the engine efficiency as a function of the system parameters are given.

\end{abstract}
\section{Introduction}\label{intro}

\begin{figure}
\begin{center}
\includegraphics[width=0.25\columnwidth]{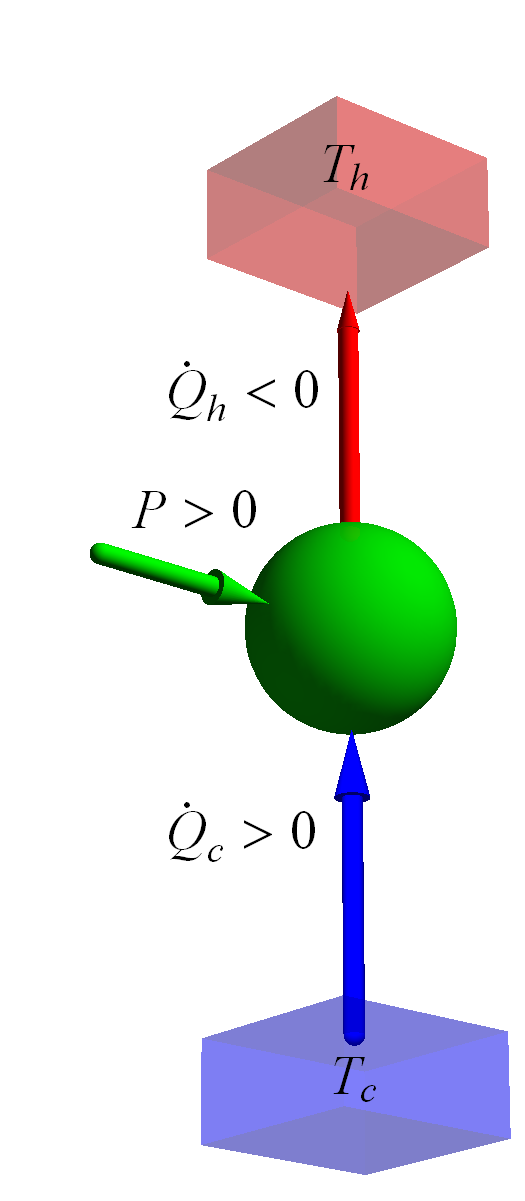} \quad \quad \quad \quad
\includegraphics[width=0.27\columnwidth]{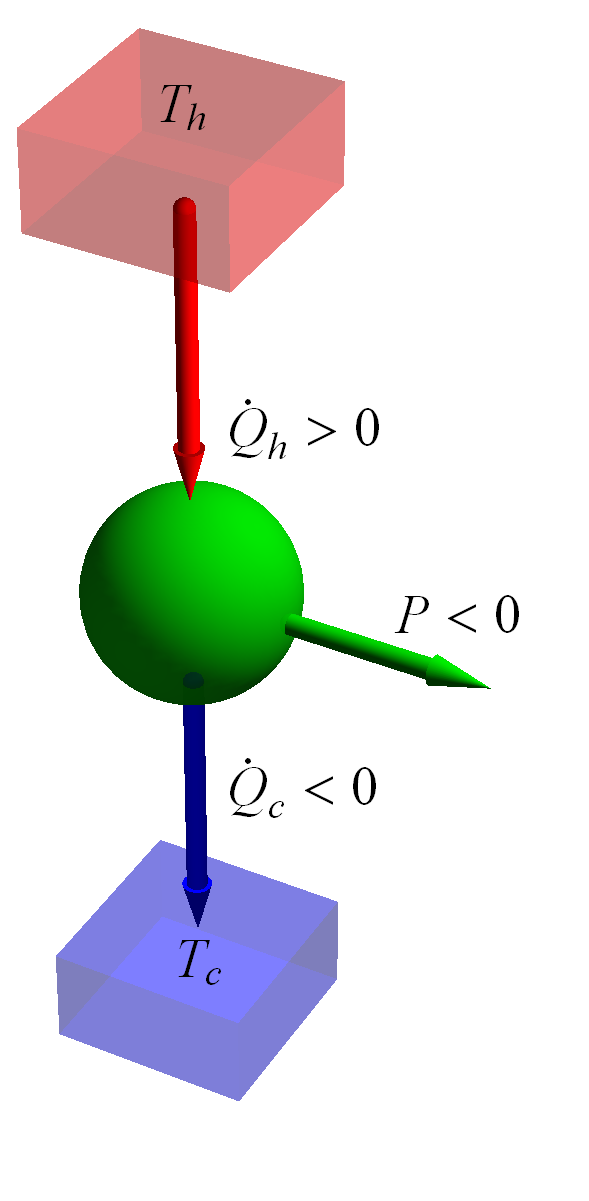}
\end{center}
\caption{Schematic representation of thermal refrigerators (left) and engines (right) coupled to a cold bath at temperature $T_c$ and a hot bath at temperature $T_h$. The heat currents $\caldot{Q}_\alpha$ ($\alpha=h,c$) and power $\cal{P}$ are defined positive when flowing towards the working system.
}
\label{Machines}       
\end{figure}

The study of the efficiency of thermal engines was one of the problems driving the development of thermodynamics \cite{Carnot1890}. Schematically, a thermal machine is represented as some working material interacting with a hot ($h$) and a cold ($c$) reservoir at temperatures $T_h$ and $T_c$, see Fig. \ref{Machines}. In the case of an engine, the system absorbs some heat at a rate $\caldot{Q}_h$ from the hot bath, casts part of it at a rate $\caldot{Q}_c$ to the cold bath and delivers useful power $\cal{P}$. Conservation of energy demands that $\caldot{Q}_h+\caldot{Q}_c+{\cal P}=0$. Similarly, the whole system acts as a refrigerator if as a result of injecting some power $\cal{P}$ in the system, some heat $\caldot{Q}_c$ is extracted from the cold bath. 
In the case of engines, a performance measure is given by the efficiency, defined as
\begin{equation}
\eta=-\frac{\cal{P}}{\caldot{Q}_h},
\label{eq:1.1}
\end{equation}
which is bounded by the Carnot efficiency $\eta_C$, $0<\eta<\eta_C=1-T_c/T_h<1$. The Carnot efficiency is only reached  in idealized models at vanishing rates, and from a practical point of view a more interesting problem is the efficiency of both classical \cite{yvon19551,novikov19571,curzon19751,vandenbroeck20051,Esposito2009} and quantum \cite{geva19921,zhou20101,abe20111,abah20121,gelbwaser20131} engines at maximum power output. In particular, the efficiency at maximum power of an optimized thermal engine in the endoreversible limit (where the only irreversible contribution is due to finite rate heat transfer effects) is given by the Curzon-Ahlborn expression \cite{yvon19551,novikov19571,curzon19751}
\begin{equation}
\eta_{CA}=1-\sqrt{1-\eta_C}.
\label{eq:1.2}
\end{equation}
Although $\eta_{CA}$ is neither an upper or lower bound for the efficiency of a general system, it describes reasonably well the behavior of actual thermal engines working with bath temperatures corresponding to low $\eta_C$. The reason of this success can be explained by considering the Taylor expansion 
\begin{equation}
\eta_{CA}=\frac{\eta_C}{2}+\frac{\eta_C^2}{8}+\cdots.
\label{eq:1.3}
\end{equation} 
The first term has been shown to be an upper bound for the efficiency at maximum power output in the linear response of engines: low $\eta_C$ implies $T_c\approx T_h$, the stationary state reached by the engine is close to thermal equilibrium, and hence the tools of linear thermodynamics can be successfully applied \cite{vandenbroeck20051}. The second term is found in the analysis of systems with strong coupled fluxes and with left-right symmetry \cite{Esposito2009}. 

A performance measure for refrigerators is the coefficient of performance (COP), defined as
\begin{equation}
\epsilon=\frac{\caldot{Q}_c}{\cal{P}},
\label{eq:1.4}
\end{equation}
which satisfies $0<\epsilon\le \epsilon_C=T_c/(T_h-T_c)<\infty$, with $\epsilon_C$ the Carnot COP. 
The optimal performance of classical and quantum refrigerators has been also extensively studied \cite{velaso19971,wang20121,allahverdyan20101}. In particular, it has been shown that the COP at maximum cooling rate $\caldot{Q}_c$ of endoreversible quantum refrigerators depends strongly on the system-bath interaction mechanism \cite{correa20131,correa20143,correa20141,correa20142}, in contrast to the simple result (\ref{eq:1.3}) found for engines. 
However, those results are not directly comparable as the refrigerator optimal performance lacks the global maximum for which Eq. (\ref{eq:1.3}) is derived in the case of engines. Besides low Carnot COP corresponds to bath temperatures far from thermodynamic equilibrium and the tools of linear thermodynamics cannot be applied. 

The purpose of this paper is to analyze the optimal performance of endoreversible thermal machines that can work either as engines or refrigerators depending on an internal control parameter. We consider this control parameter as the only optimization variable in our system. 
In section \ref{section 2} we motivate our study using a simple quantum model, the three-level maser. In this model the larger normalized optimal performances, $\eta/\eta_C$ and $\epsilon/\epsilon_C$, are reached in the regime of low Carnot efficiency and low Carnot COP respectively, together with small thermodynamic forces. This will be the regime of interest in our analysis.
In section \ref{section 3} we present a generic model of endoreversible thermal machine. This model has been previously used to analyze the optimal COP in Ref. \cite{correa20142}. Here we extend this analysis to engines and discuss how the dependence of the thermodynamic flux on the forces determines the optimal performances as functions of the Carnot efficiency and COP. In particular, we obtain explicit analytical expressions for the engine optimal efficiency.  
In section \ref{section 4} we illustrate the results using the three-level maser and the main conclusions are drawn in section \ref{section 5}.


\section{The three-level maser}\label{section 2}

The three-level maser \cite{scovil19591,kosloff20131,kosloff20132} is probably the simplest model of endoreversible thermal machine.  
The system has three levels with Bohr frequencies $\omega_c$, $\omega_h$, and $\Omega=\omega_h-\omega_c$. It is periodically driven by an external field tuned to frequency $\Omega$ and weakly coupled to external unstructured bosonic baths at fixed temperatures $T_c$ and $T_h$. The cold and hot baths address the transitions $\omega_c$ and $\omega_h$, respectively. In the weak driving limit the stationary heat currents and power can be consistently obtained \cite{correa20143,kosloff20132}
\begin{eqnarray}
\caldot{Q}_c &=& \omega_c\, \mathcal{I}\,, \cr
\caldot{Q}_h &=& -\omega_h \,\mathcal{I}\,, \cr
\cal{P} &=& -\caldot{Q}_h-\caldot{Q}_c\,=\,-(\omega_c-\omega_h)\, \mathcal{I},
\label{eq:}
\end{eqnarray}
where the flux $\mathcal{I}$ is given in terms of the relaxation rates of the baths as 
\begin{equation}
\mathcal{I}=\frac{\Gamma_h \Gamma_c \left( e^{-\omega_c/T_c}- e^{-\omega_h/T_h} \right)}
{\Gamma_h(1+2 e^{-\omega_h/T_h} )+\Gamma_c(1+2 e^{-\omega_c/T_c})}.
\label{I1}
\end{equation}
The heat current $\caldot{Q}_c$  will be referred as \emph{cooling rate} when the machine operates as a refrigerator, and $\cal{P}$ simply as \emph{power}. 

The rates in (\ref{I1}) for a bosonic bath are given by \cite{breuer2002a} 
\begin{equation}
\Gamma_{\alpha}\,=\,\gamma_\alpha \omega_\alpha^{d_\alpha}\left[1+N(\omega_\alpha)\right]\,,
\label{eq:}
\end{equation}
with $N(\omega_a)=\left( e^{\omega_a/T_a}-1\right)^{-1}\,$, $d_\alpha$ the physical dimensionality of the bath, and $\alpha=h,c$.

\begin{figure}[htp]%
\begin{center}
\includegraphics[width=0.75\columnwidth]{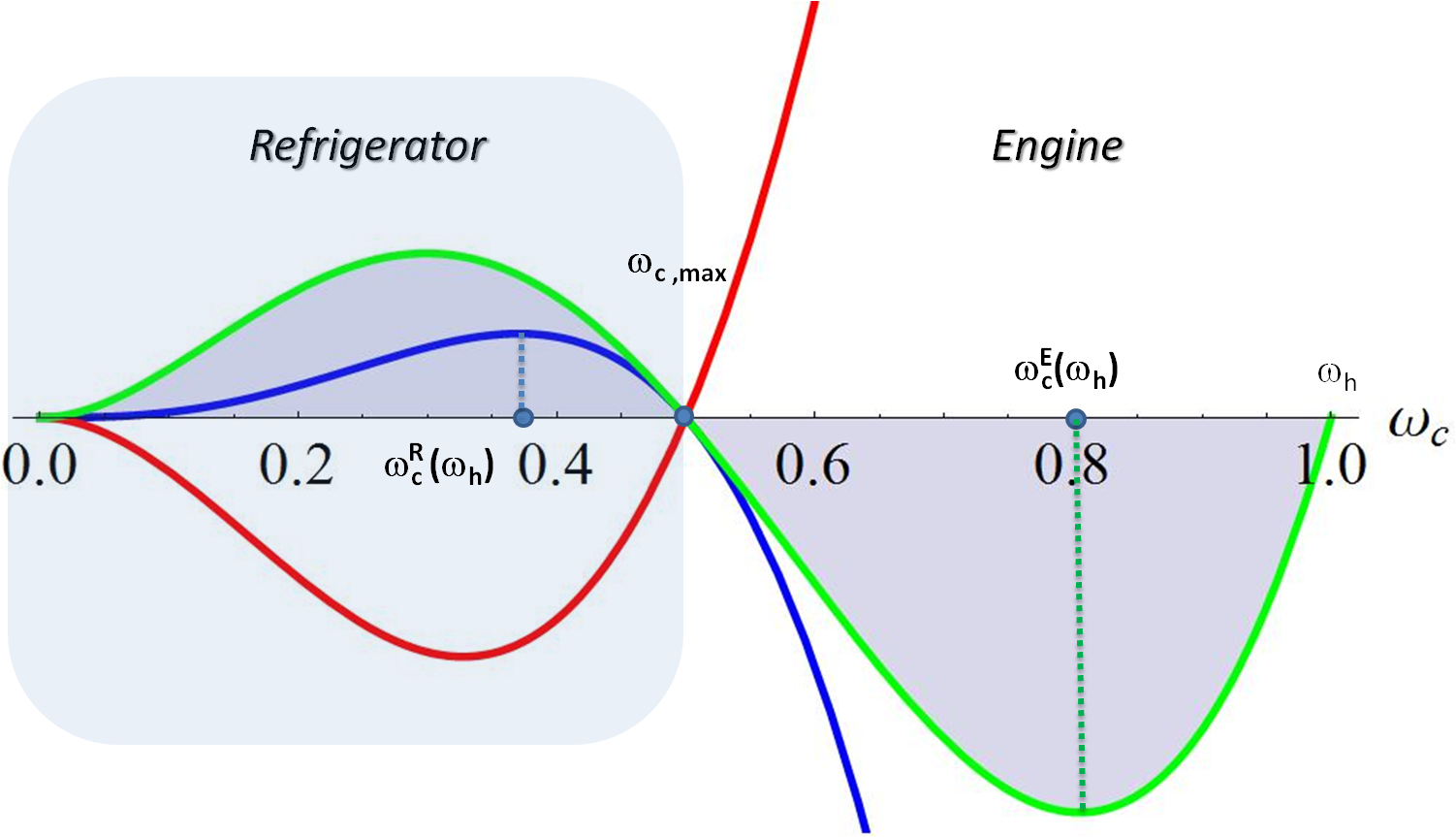}%
\end{center}
\caption{Schematic representation of the cold energy current ${\caldot{Q}_c}$ (blue line), hot current ${\caldot{Q}_h}$ (red line) and power ${\cal{P}}=-({\caldot{Q}_c}+{\caldot{Q}_h})$ (green line) in the three-level maser as functions of the cold transition frequency $\omega_c$. The system works as a refrigerator for cold frequencies in the interval $0\leq \omega_c \leq \omega_{c,max}=\omega_h T_c/T_h$, and as an engine for $\omega_{c,max}\leq \omega_c \leq \omega_h$. The optimal cold frequencies for maximum cooling rate $\omega_{c}^R$ and maximum power output $\omega_{c}^E$ are determined by the model parameters, in particular $\omega_h$.} 
\label{QPfig1}%
\end{figure}

We assume some degree of control only over the system frequencies, $\omega_c$ and $\omega_h$, and the bath temperatures, $T_c$ and $T_h$. The three-level maser can operate either as a refrigerator or as an engine depending on $\omega_c$, which in the following will be considered the optimization variable. As depicted in Fig. \ref{QPfig1}, the cooling rate has a maximum at some frequency $\omega_c^R$ in the cooling window $0<\omega_c^R<\omega_{c,max}$, and the power output has its maximum at a given frequency $\omega_c^E$ such that $\omega_{c,max} \le \omega_c^E \le \omega_h$. When $\omega_c=\omega_{c,max}$ the machine reaches the Carnot efficiency 
\begin{equation}
\eta({\omega_{c,max}})=\eta_C, \quad \epsilon(\omega_{c,max})=\epsilon_C\,.
\label{eq:}
\end{equation}
but at zero power and heat currents.

To analyze the optimization of refrigerators and engines we introduce the thermodynamic forces $x_c=\omega_c/T_c$ and $x_h=\omega_h/T_h$, related to the cold and hot bath temperatures respectively. The heat currents and the power are rewritten as
\begin{eqnarray}
\caldot{Q}_c &=& T_c\, x_c\, \mathcal{I}\,, \cr
\caldot{Q}_h &=& -T_h\, x_h\, \mathcal{I}\,, \cr
\cal{P} &=& -\caldot{Q}_h-\caldot{Q}_c=(T_h x_h-T_c x_c)\, \mathcal{I}.
\label{eq:flows}
\end{eqnarray}
The entropy production is a bilinear form of the forces $x_h$ and $x_c$ and the flux $\mathcal{I}$
\begin{equation}
\caldot{S}\,=\,-\frac{\caldot{Q}_h}{T_h}-\frac{\caldot{Q}_c}{T_c}\,=\,(x_h-x_c)\,\mathcal{I},
\end{equation}
and the efficiency and COP of the machine are
\begin{eqnarray}
\eta &=& 1-(1-\eta_C)x_c/x_h, \cr\cr
\epsilon &=& \frac{\epsilon_C}{(1+\epsilon_C)x_h/x_c-\epsilon_C}.
\label{eq:efficiency}
\end{eqnarray}
According to the previous definitions it follows that
\begin{equation}
   \mathcal{I} = \left\{
     \begin{array}{lcl}
       >0 & : x_c<x_h\,, & {\rm (Refrigerator)}\\
       0 &  : x_c=x_h\,, & \\
       <0 &  : x_c>x_h\,, & {\rm (Engine)}.\\
     \end{array}
   \right.
\label{eq:current}
\end{equation}

The optimization variable will be then $x_c$. The maximum cooling rate is reached for an optimal force $x_c^R=\omega_c^R/T_c<x_h$ and the maximum power output for $x_c^E=\omega_c^E/T_c>x_h$.
Finally we explicitly introduce the Carnot efficiency and COP into the discussion replacing $T_c$ by
\begin{equation}
T_c \longmapsto(1-\eta_C)T_h,
\label{eq: }
\end{equation}
in the case of engines, and by
\begin{equation}
T_c \longmapsto T_h \epsilon_C/(1+\epsilon_C),
\label{eq: Tc COP}
\end{equation}
when dealing with refrigerators.


\subsection{Optimal cooling rate and power output}

The optimal force can be obtained from the equation
\begin{equation}
\frac{\partial \caldot{Q}_c}{\partial x_c}=\mathcal{I}+x_c \frac{\partial \mathcal{I}}{\partial x_c}=0,
\label{eq:exref}
\end{equation}
expressing the extreme condition when working as a refrigerator, and by
\begin{equation}
\frac{\partial \mathcal{P}}{\partial x_c}=-(1-\eta_C)\mathcal{I}+\left(x_h-\left(1-\eta_C\right)x_c\right)\frac{\partial \mathcal{I}}{\partial x_c}=0,
\label{eq:exeng}
\end{equation}
when operating as an engine.

\begin{figure}[t]%
\begin{center}
\includegraphics[width=0.45\columnwidth]{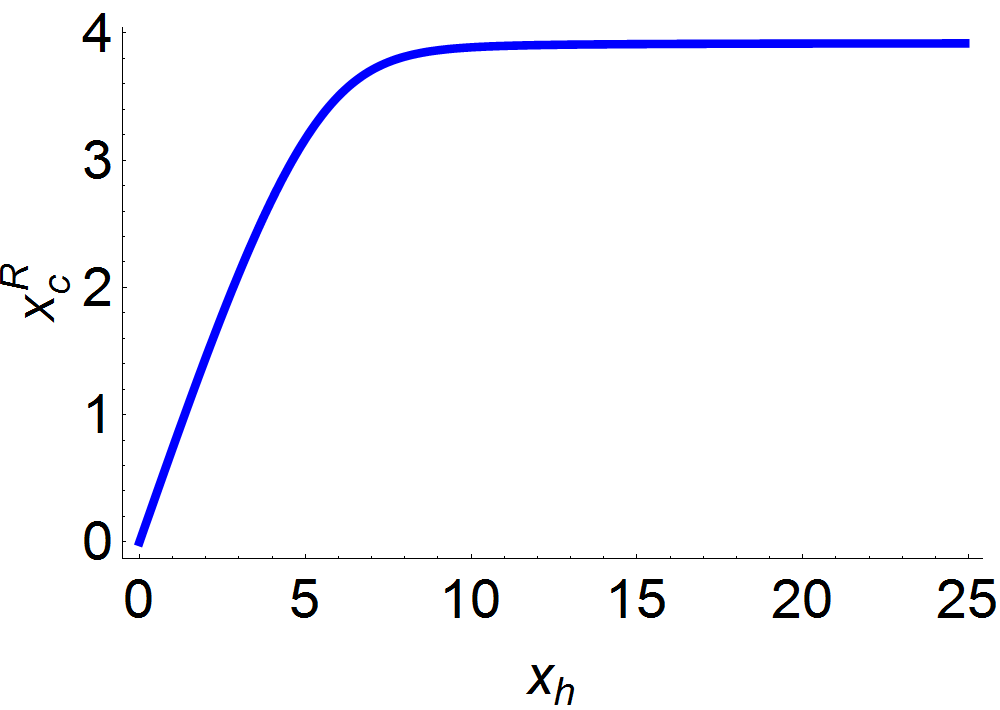}%
\includegraphics[width=0.45\columnwidth]{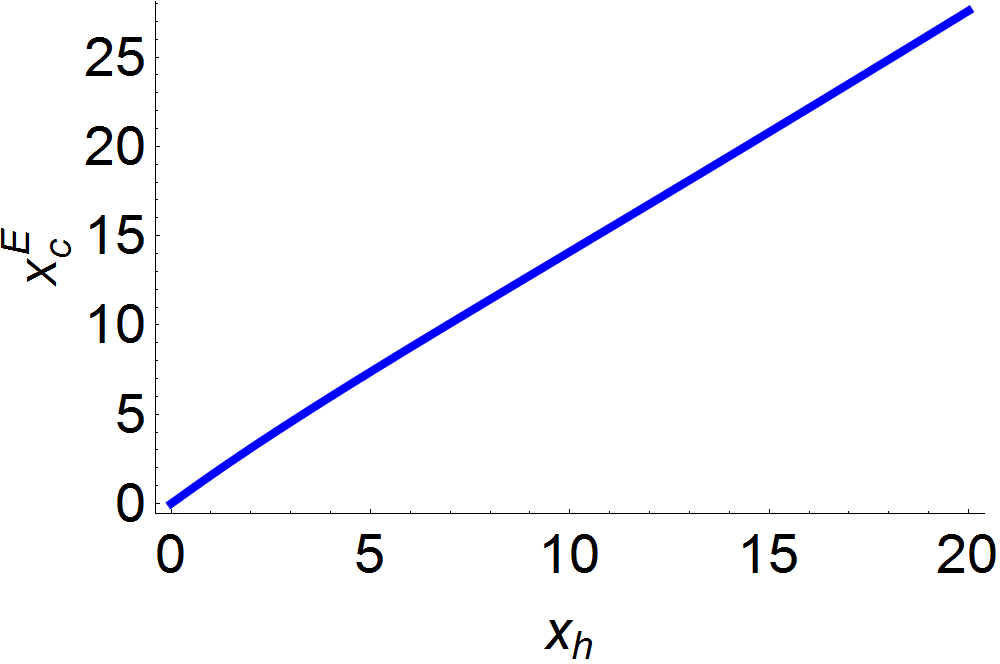}
\end{center}
\caption{Optimal force $x^R_c$ (left) and $x^E_c$ (right) as a function of 
$x_h$ for $T_c=5$, $T_h=10$ (in arbitrary units), $d_c=d_h=3$ and $\gamma_c/\gamma_h=1$.}
\label{wc}%
\end{figure}

The solution of the optimization problem in the $x_c$ variable for fixed $T_c$ and $T_h$ gives a function $x_{c}^{(R,E)}$ of $x_h$, where the $R,E$ refers to refrigerators and engines respectively. In both cases, such function increases monotonically with $x_h$, but for refrigerators it saturates to a fixed value, see Fig. \ref{wc}. For $d_c=d_h=d$ this value is given by 
\begin{equation}
x_{c}^{R}(x_h\to \infty)=\frac{5\left(d+1+\mathcal{W}\left(\left(-d-1\right)e^{-d-1}\right)\right)}{T_c}\,, 
\end{equation}
where $\mathcal{W}(x)$ the Lambert function \cite{Lambert_Fuction}. 

\begin{figure}[t]%
\begin{center}
\includegraphics[width=0.45\columnwidth]{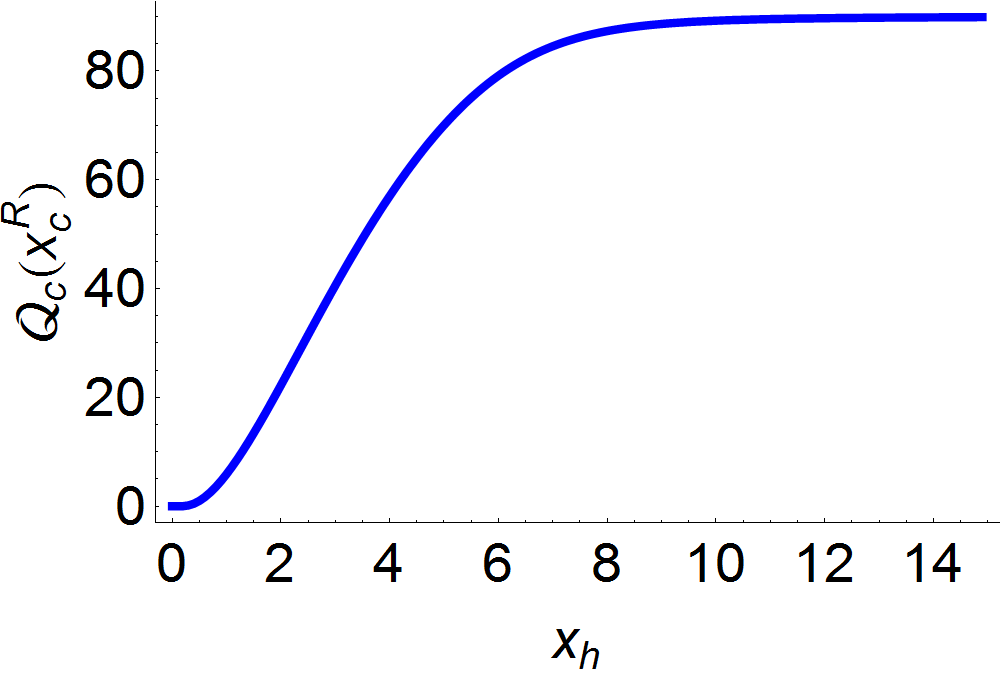}%
\includegraphics[width=0.45\columnwidth]{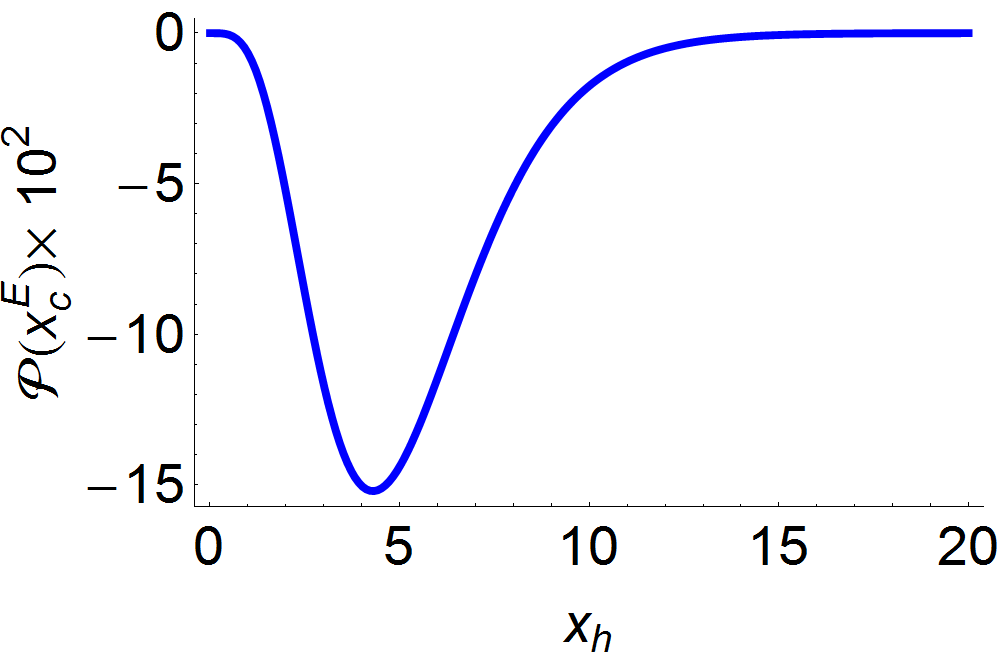}
\end{center}
\caption{Optimal cooling rate (left) and optimal power output 
(right) as a function of $x_h$ for $T_c=5$, $T_h=10$ (arbitrary units), $d_c=d_h=3$ and $\gamma_c/\gamma_h=1$.}%
\label{QcP}%
\end{figure}

The optimal cooling rate also increases monotonically with $x_h$ until saturation, as shown in Fig. \ref{QcP}. In contrast, the optimal power output has a minimum value at a given $x_h$ and a further optimization in this variable is possible. This additional optimization is considered for example in Ref. \cite{Esposito2009}. To keep a fair comparison between refrigerators and engines we focus on a partial optimization in the control variable $x_c$.

\begin{figure}[t]%
\begin{center}
\includegraphics[width=0.45\columnwidth]{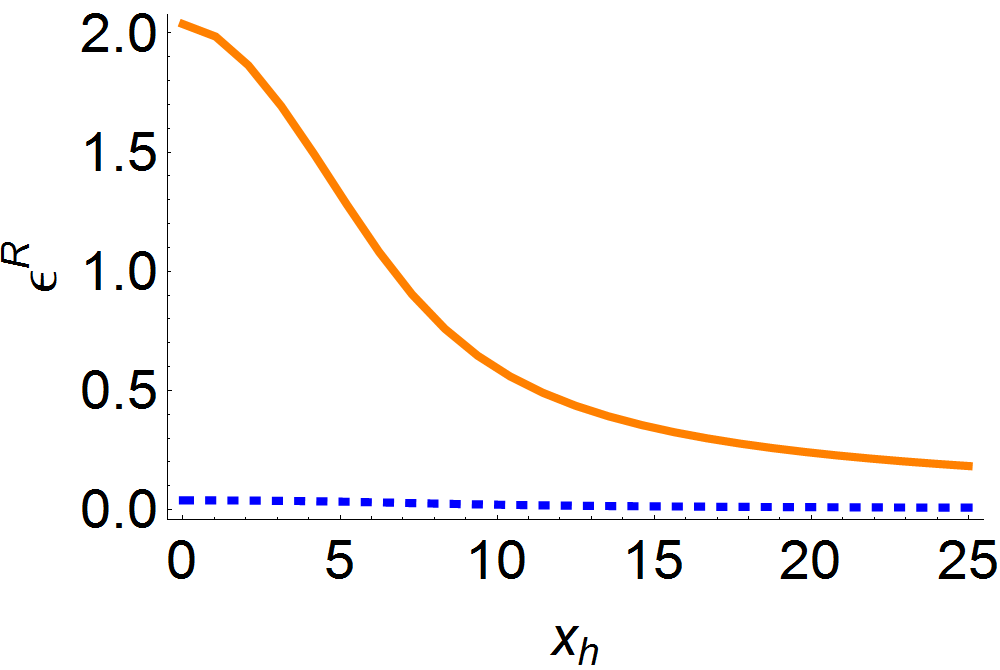}%
\includegraphics[width=0.45\columnwidth]{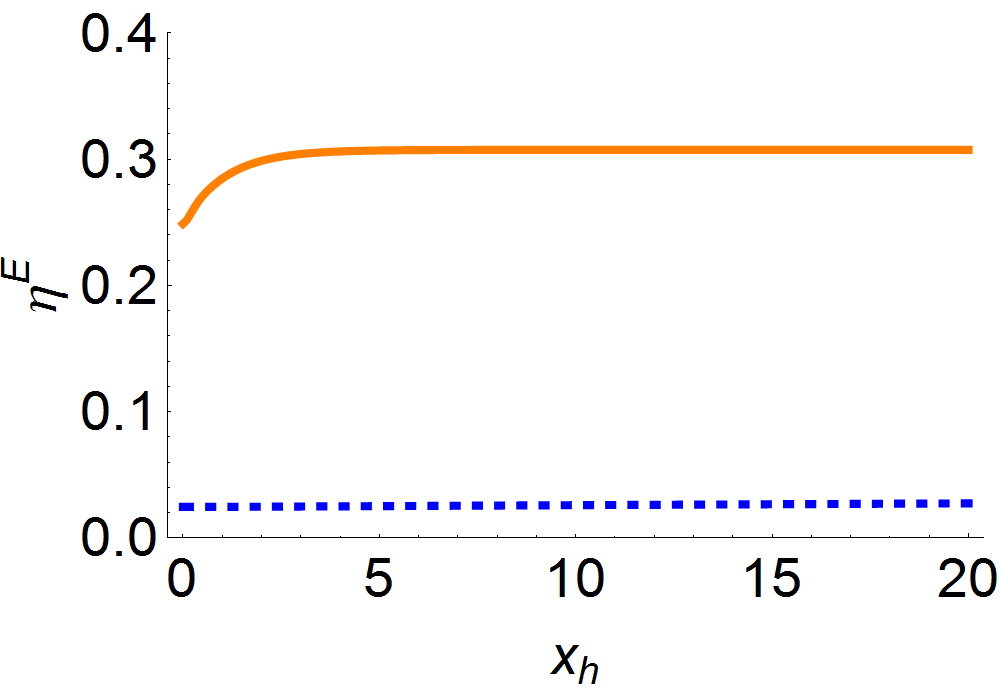}
\end{center}
\caption{COP and efficiency at the optimal point as a function of force $x_h$ for the three-level refrigerator (left) and engine (right). The Carnot COPs are $\epsilon_C=0.05$ (blue dashed line) and $\epsilon_C=19$ (orange solid line). The Carnot efficiencies are $\eta_C=0.05$ (blue dashed line) and $\eta_C=0.95$ (orange solid line). We set $\omega_h=1$, $T_h=10$ (arbitrary units), $d_c=d_h=3$ and $\gamma_c/\gamma_h=1$.}
\label{opteff}%
\end{figure}

Next we discuss the efficiency ($\eta^E$) and COP ($\epsilon^R$) for the optimal force $x_{c}^{(R,E)}$. In Fig. \ref{opteff} we show them as a function of the force $x_h$ for two cases, low and large Carnot efficiency and COP. 

\begin{figure}[t]%
\begin{center}
\includegraphics[width=0.45\columnwidth]{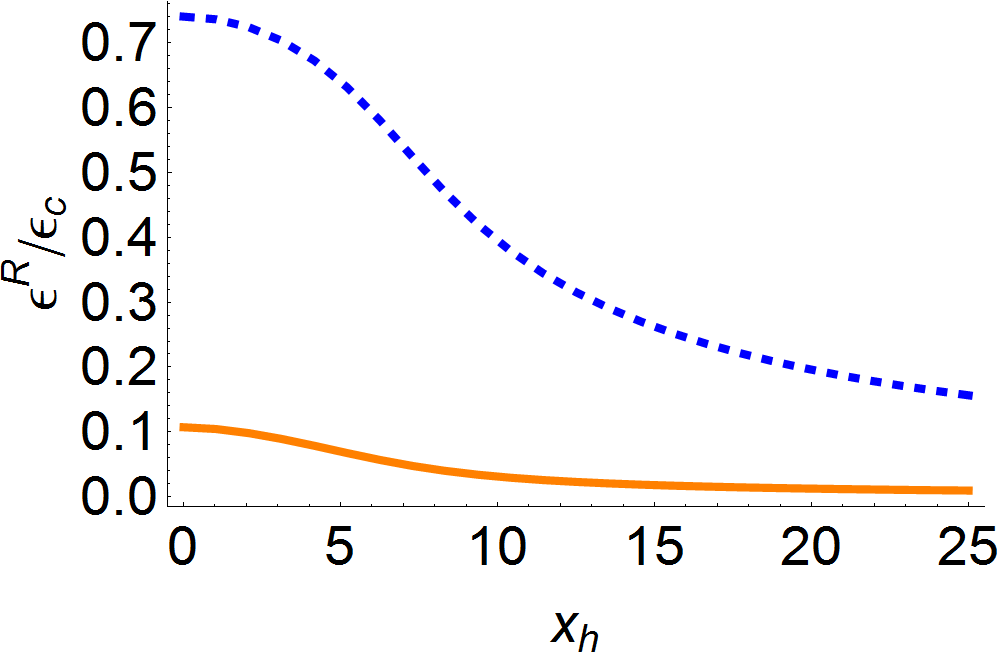}%
\includegraphics[width=0.45\columnwidth]{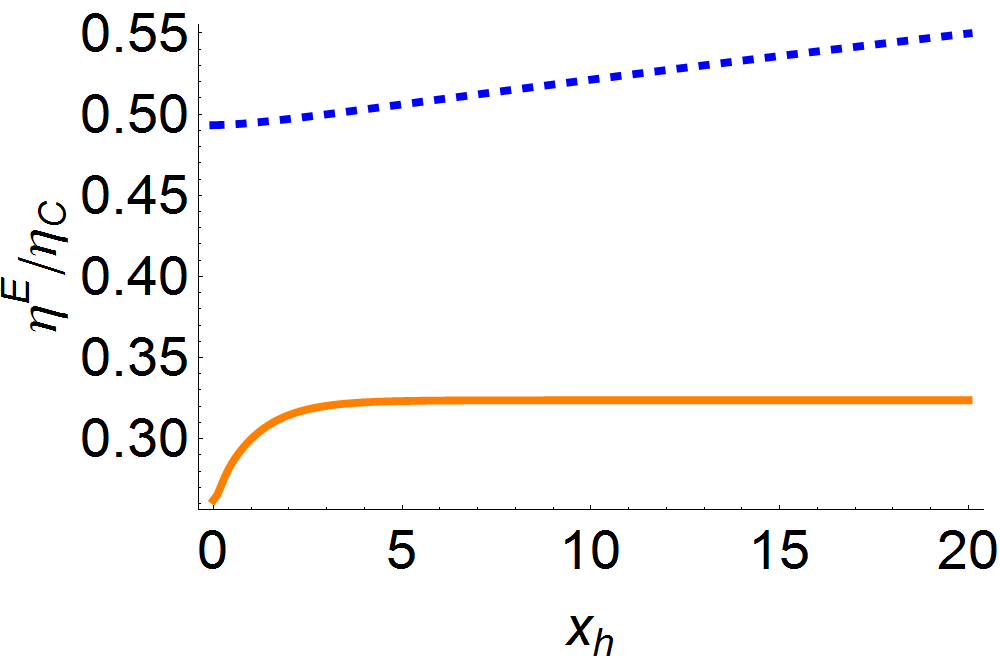}
\end{center}
\caption{Normalized performances obtained from figure \ref{opteff}.}
\label{opteffC}%
\end{figure}

A different scenario emerges in Fig. \ref{opteffC} when one looks at the normalized performances, defined as $\epsilon^R/\epsilon_C$ for refrigerators and $\eta^E/\eta_C$ for engines. Now, the less efficient machines have the largest normalized performances. In the case of refrigerators, although the cooling rate increases with $x_h$, the COP, both absolute and normalized, decreases. Besides, the larger values of the normalized COP and efficiency correspond to bath temperatures leading to low $\epsilon_C$ and $\eta_C$ repectively. These results set the conditions of interest for our comparison: small thermodynamic forces and low $\epsilon_C$ and $\eta_C$.

\begin{figure}[t]%
\begin{center}
\includegraphics[width=0.45\columnwidth]{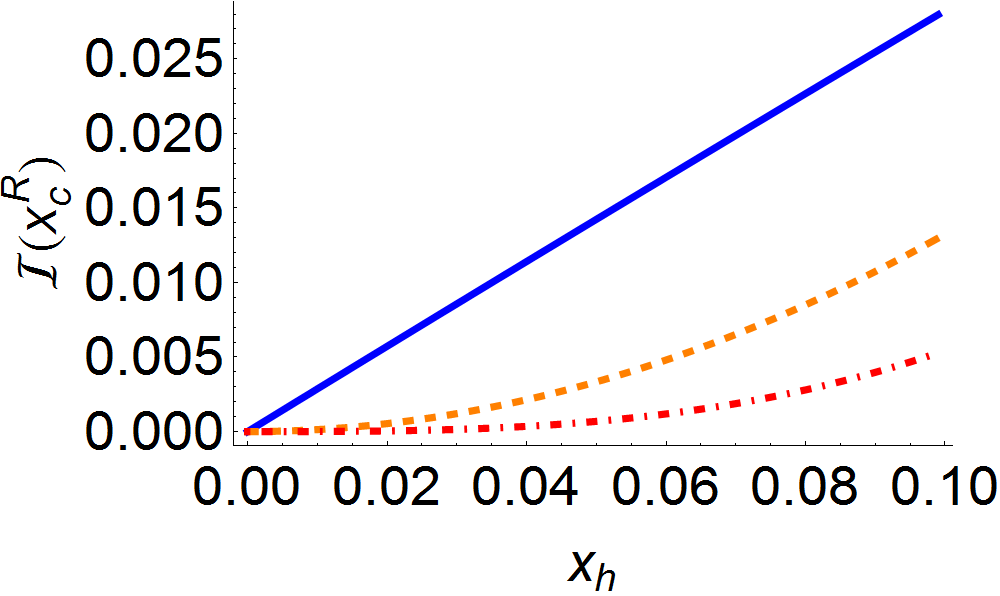}%
\includegraphics[width=0.45\columnwidth]{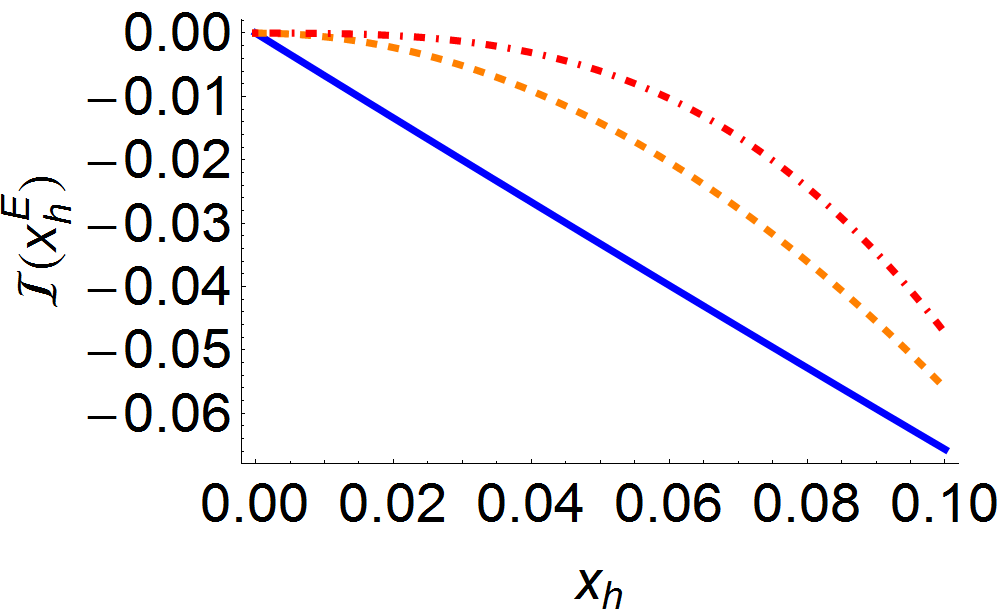}
\end{center}
\caption{Flux $\mathcal{I}$ at the optimal point $x_c^{(R,E)}$ as a function of $x_h$ for the three-level refrigerator (left) and engine (right) setting the bath dimensions $d_c=d_h=1$ (blue solid line), 2 (orange dashed line) and 3 (red dashed-dotted line). The other parameters are $T_c=5$, $T_h=10$ (arbitrary units) and $\gamma_c/\gamma_h=1$.}%
\label{I}%
\end{figure}

The efficiency (\ref{eq:efficiency}) at the optimal point is determined by $x_c^{(R,E)}$, which in turns depends on the explicit form of $\mathcal{I}$ through Eqs. (\ref{eq:exref}) and (\ref{eq:exeng}). Hence, an important feature of the model is the relation between the flux $\mathcal{I}$ and the thermodynamical forces. Figure \ref{I} shows the flux at the optimal point for small $x_h$. A different behavior is found depending on the physical dimensionality of the bath. In next section we show how such behavior determines the normalized performance in each case.

	
\section{Performance in the regime of small forces}\label{section 3}

In this section we study a generic model of endoreversible machine in the regime of small thermodynamic forces. In this model the heat currents and power are given by Eqs. (\ref{eq:flows}) and (\ref{eq:current}). The particular nature of a given classical or quantum machine will be then determined by the relation between the flux $\mathcal{I}$ and the system parameters. However, in our discussion only two very general assumptions about its dependence on the thermodynamic forces are needed: when $x_h$ vanishes (a) the optimal point $x_c^R$ for the refrigerator and (b) the optimal point $x_h^E$ for the engine go to zero, as found in the three-level maser, see Fig. \ref{wc}. 

The analysis of the optimal performance follows the study in Ref. \cite{correa20142} for refrigerators. Using the previous assumptions, the optimal value $x_{c}^{(R,E)}$ can be approximated by a power series in $x_h$,
\begin{equation}
x_{c}^{(R,E)}\,=\,C_1^{(R,E)} x_h+C_2^{(R,E)} x^2_h+\cdots
\label{taylor1}
\end{equation}
In the case of refrigerators and for small $x_h$ we can keep the first term, $x_{c}^{R} \approx C_1^{R} x_h$. The coefficient $C_1^{R}$ will be a function of all parameters of the problem, satisfying
\begin{equation}
C_1^{R}(T_h,\epsilon_C,\Gamma_c,\Gamma_h,...)\le 1.
\label{c1r}
\end{equation}
Moreover, it determines the COP at maximum cooling rate as 
\begin{equation}
\epsilon^{R}=\frac{\epsilon_C}{(\epsilon_C+1)x_h/x_{c}^{R}-\epsilon_C}=\frac{C_1^{R} \epsilon_C}{(1-C_1^{R})\epsilon_C+1}.
\label{eq:}
\end{equation}

The previous discussion can be applied also for engines. Now $x_{c}^{E} \approx C_1^{E} x_h$ with $C_1^{E}\ge 1$, in terms of which the efficiency at maximum output power is
\begin{equation}
\eta^{E}=1-(1-\eta_C)x_{c}^{E}/x_h=1-(1-\eta_C)C_1^{E}.
\label{eq:}
\end{equation}

In the regime of small forces, the coefficients $C_1^{(R,E)}$ will be determined by the first non-zero term of the Taylor expansion of the current $\mathcal{I}$
\begin{eqnarray}\label{Itaylor}
\mathcal{I}(x_c,x_h) &=& \sum_{i=c,h}\mathcal{I}_{i}(0,0) x_i \,+\,
\frac{1}{2!}\sum_{i,j=c,h}\mathcal{I}_{ij}(0,0)x_i x_j \,+\, \cr
&+& \frac{1}{3!}\sum_{i,j,k=c,h}\mathcal{I}_{ijk}(0,0)x_i x_j x_k \,+\, \cdots\,,
\end{eqnarray} 
as discussed below.


\subsection{Linear term}

Let us assume that the first non zero term in (\ref{Itaylor}) is the linear term. Then 
\begin{equation}
\mathcal{I}\approx \mathcal{I}_0(x_h-x_c) 
\label{eq:}
\end{equation}
because $\mathcal{I}(x_c=x_h)=0$. As a consequence, it follows that
\begin{eqnarray}
C_1^{R} &=& \frac{1}{2}\,, \cr 
C_1^{E} &=& \frac{2-\eta_C}{2(1-\eta_C)}.
\label{eq:}
\end{eqnarray}
With these coefficients the normalized COP and efficiency at the optimal forces are given by
\begin{equation}
\frac{\epsilon^R}{\epsilon_C}=\frac{1}{2+\epsilon_C}\,,
\label{eq:elR}
\end{equation}
and
\begin{equation}
\frac{\eta^E}{\eta_C}=\frac{1}{2}\,.
\label{eq:elE}
\end{equation}

\subsection{Higher orders}\label{higher_order}

When the first order is zero, higher orders in the expansion (\ref{Itaylor}) must be considered. For simplicity, let us assume the particular form
\begin{equation}
\mathcal{I}= \mathcal{I}_0 x_c^{d-1}(x_h-x_c),
\label{eq:simpleflux}
\end{equation}
of the current for small forces, where $d$ is a model parameter. With this choice, the first non-zero term in the expansion will be the $d$-term. For example, when $d=1$, the linear case is recovered. Applying the same procedure as before, it follows that
\begin{equation}
C_1^{R}=\frac{d}{d+1},
\label{eq:}
\end{equation}
what leads to the normalized COP at the optimal force \cite{correa20143}
\begin{equation}
\frac{\epsilon^R}{\epsilon_C}=\frac{d}{d+1+\epsilon_C}.
\label{eff_refrigerators}
\end{equation}

In the case of engines, the same procedure gives
\begin{equation}
C_1^{E}=\frac{d(2-\eta_C)+\sqrt{d^2 \eta_C^2-4 \eta_C+4}}{2(d+1)(1-\eta_C)}
\label{eq:}
\end{equation}
and the normalized efficiency at maximum power
\begin{equation}
\frac{\eta^E}{\eta_C}=\frac{2+d\,\eta_C-\sqrt{d^2 \eta_C^2-4 \eta_C+4}}{2(d+1)\eta_C}.
\label{eff_engines}
\end{equation}
In the limit of vanishing Carnot efficiency $\eta^E \rightarrow \eta_C/2$.

We have shown that the normalized performances are determined by the flux $\mathcal{I}$ and therefore by details of the model such as the coupling between system and baths. In the linear case $\epsilon^R/\epsilon_C$ and $\eta^E/\eta_C$ tend to $1/2$ in the limit of vanishing Carnot efficiency and COP. However, for higher orders the normalized COP could saturate at different values, see Eq. (\ref{eff_refrigerators}). In contrast, $\eta^E/\eta_C$ always saturates at $1/2$ in the limit of vanishing Carnot efficiency.


\section{Example: the three-level maser in the regime of high temperatures.}\label{section 4}

In the regime of high temperatures, in which the thermodynamic forces $x_c,x_h\ll 1$, the flux (\ref{I1}) is given by
\begin{equation}
\mathcal{I} \approx \frac{\Gamma_c (x_h-x_c)}{3(1+\Gamma_c/\Gamma_h)}\,,
\label{eq:aflux}
\end{equation}
with
\begin{equation}
\frac{\Gamma_c}{\Gamma_h}=\frac{\gamma_c}{\gamma_h}\,\frac{x_c^{d_c-1}}{x_h^{d_h-1}}\,\frac{T_c^{d_c}}{T_h^{d_h}}\,.
\label{eq:}
\end{equation}
Hence the dependence of the denominator of Eq. (\ref{eq:aflux}) on $x_c$ and $x_h$ can be neglected when: (a) $d_h=d_c=1$, (b) $\gamma_c\ll\gamma_h$, or (c) $\epsilon_C \ll 1$, see Eq. (\ref{eq: Tc COP}). In such cases
\begin{equation}
\mathcal{I}\propto x_c^{d_c-1}(x_h-x_c),
\label{eq:simplefluxb}
\end{equation}
which matches Eq. (\ref{eq:simpleflux}). In particular, when $\epsilon_C\longrightarrow 0$ Eq. (\ref{eff_refrigerators}) gives
\begin{equation}
\frac{\epsilon^R}{\epsilon_C}=\frac{d_c}{d_c+1},
\end{equation}
which is the maximum normalized efficiency for optimized quantum refrigerators coupled to bosonic cold baths of physical dimensionality $d_c$ \cite{correa20131,correa20143,correa20141,correa20142}. 

\begin{figure}[t]%
\begin{center}
\includegraphics[width=0.45\columnwidth]{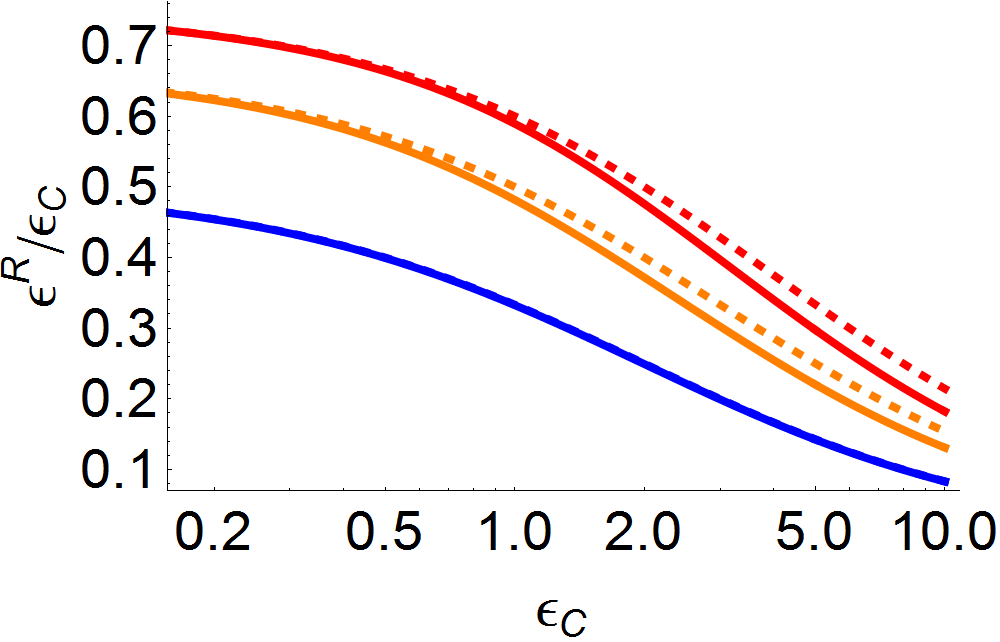}%
\includegraphics[width=0.45\columnwidth]{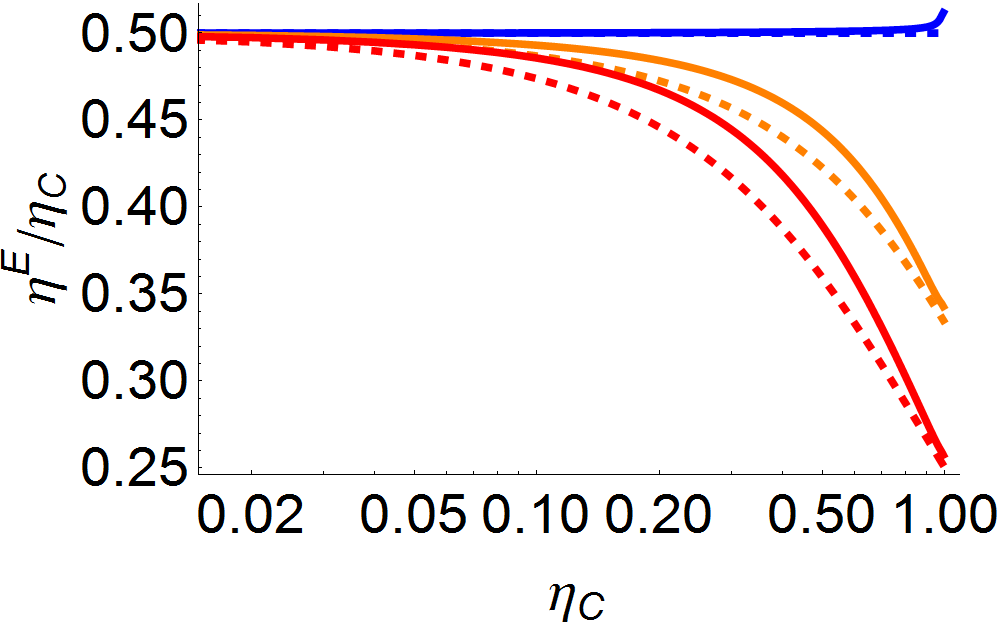}
\end{center}
\caption{Normalized optimal performance  for three-level refrigerators (left) and engines (right) for $d_c=d_h=1$ (blue solid line), $2$ (orange solid line), $3$ (red solid line) and $\gamma_c/\gamma_h=1$. The dashed lines are the results of Eqs. (\ref{eff_refrigerators}) and (\ref{eff_engines}) in each case. The parameters used are $\omega_h=1$, $T_c=5$ and $T_h=10$ (arbitrary units).}%
\label{effsym}%
\end{figure}

\begin{figure}[t]%
\begin{center}
\includegraphics[width=0.45\columnwidth]{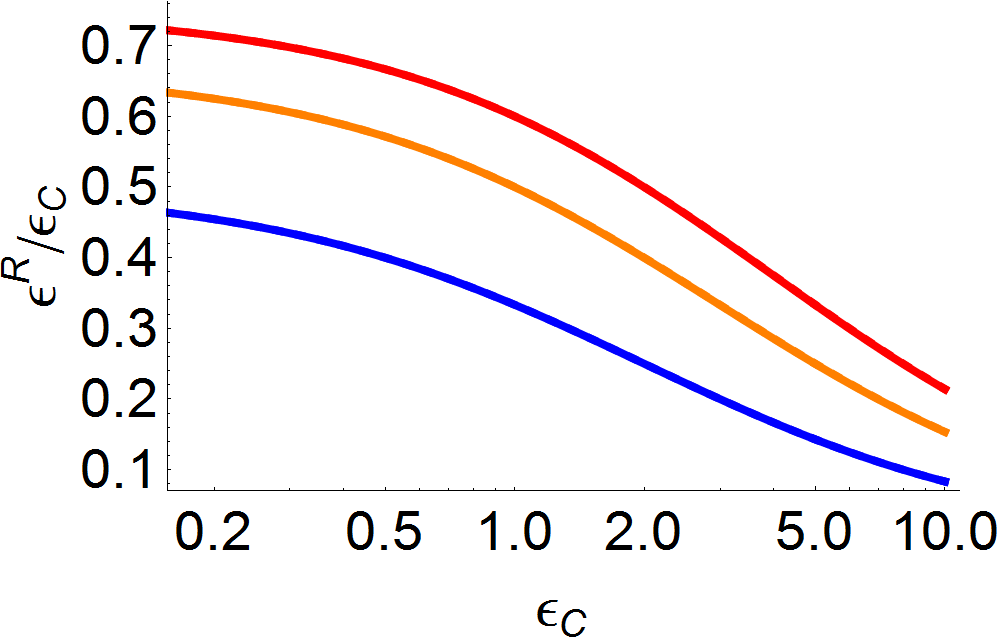}%
\includegraphics[width=0.45\columnwidth]{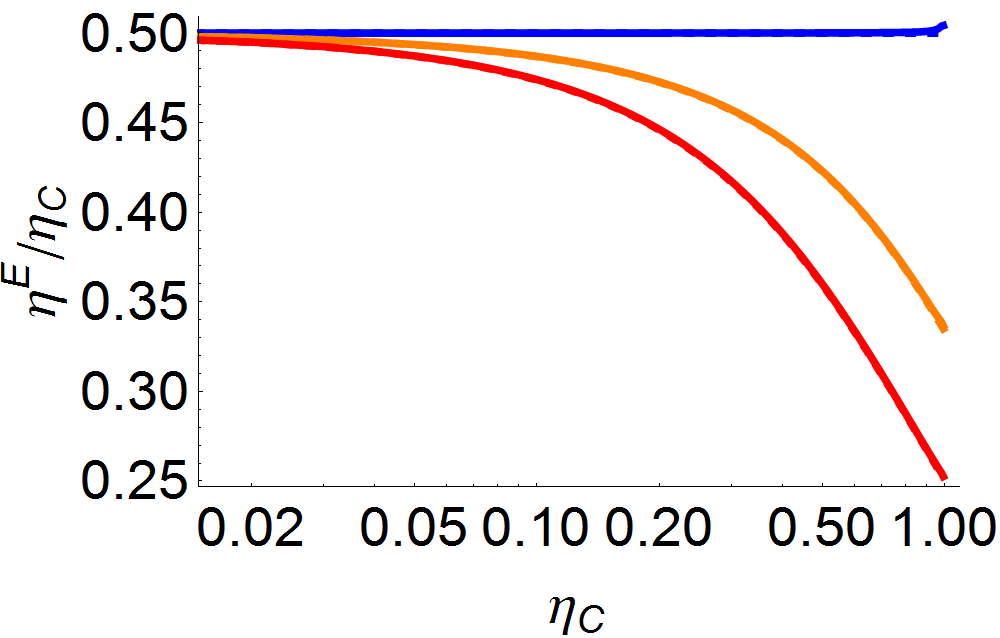}
\end{center}
\caption{As in figure \ref{effsym} but for $\gamma_c/\gamma_h=0.01$.}%
\label{effnonsym}%
\end{figure}

Figure \ref{effsym} shows $\epsilon^R/\epsilon_C$ and $\eta^E/\eta_C$ for equal coupling strengths, $\gamma_c/\gamma_h=1$. In this case the equations derived in section \ref{section 3} are good approximations for the normalized efficiencies of the thermal machine. The agreement with Eqs. (\ref{eff_refrigerators}) and (\ref{eff_engines}) improves when $\gamma_c/\gamma_h<1$, as shown in figure \ref{effnonsym}. 

\section{Conclusions}\label{section 5}

In this paper we have studied the normalized performances of optimized endoreversible thermal machines working either as refrigerators or as engines depending on the values of the thermodynamic force related to the cold bath temperature, which has been chosen as the control parameter. The optimization has been performed with respect to this parameter to obtain the maximum cooling rate or power output. This is the main difference with previous studies on the efficiency at maximum power output, see for instance \cite{yvon19551,novikov19571,curzon19751,vandenbroeck20051,Esposito2009}, where an optimization involving all thermodynamic forces is considered. We have focused on the regime of small thermodynamic forces and low $\epsilon_C$ and $\eta_C$. We must emphasize that bath temperatures leading to low $\epsilon_C$ or $\eta_C$ correspond to very different physical situations in each case: very far from equilibrium for refrigerators, as $T_c \ll T_h$, and  very close to equilibrium for engines. However, by considering an optimization with respect to a single parameter and low Carnot efficiency and COP, the optimal performance of the thermal machine can be analyzed for both operating modes under the same conditions. 

Considering very general assumptions about the endoreversible machine, we have determined the normalized performance as a function of the Carnot efficiency and COP. For vanishing COP, $\epsilon^R/\epsilon_C$ saturates at different values depending on which is the first non-zero term in the Taylor expansion of the flux. The value $1/2$ is reached only when the flux is linear in the thermodynamic forces \cite{correa20142}. In contrast, we have found that for engines $\eta^E/\eta_C$ always saturates at $1/2$ in spite of the dependence of the optimal efficiency on the system parameters. Although the results have been illustrated using a quantum endoreversible model, the three-level maser, our analysis applies to any system, classical or quantum, given some knowledge about the dependence of the flux on the thermodynamic forces. 

We have focused on endoreversible systems but additional sources of irreversibility can be considered to model more realistic machines. Their effect in the system efficiency and COP has been studied for example in Ref. \cite{wang20121,esposito2010a,correa2015}. The question whether those additional sources of irreversibility have a different influence on the optimal performance when considering refrigerators or engines will be the subject of future work.


\begin{acknowledgements}
D. Alonso acknowledges the warm hospitality and support of the organizing committee of the \emph{Quantum Information and Thermodynamics} workshop held at S$\tilde{a}$o Carlos in February 2015 and to Prof. Lucas C. C\'eleri for his kind assistance. We thank R. Kosloff, R. Uzdin, M Esposito, A del Campo and I. de Vega for fruitful discussions at S$\tilde{a}$o Carlos. 
Financial support from Spanish MINECO (FIS2013-40627-P and FIS2013-41352-P), COST Action MP1209 and EU Collaborative Project TherMiQ (Grant Agreement 618074) is gratefully acknowledged.
\end{acknowledgements}



\end{document}